\documentclass[journal]{IEEEtran}
\usepackage{cite}
\usepackage{threeparttable}
\usepackage{graphicx}
\usepackage{picinpar}
\usepackage{amssymb}
\usepackage[cmex10]{amsmath}
\usepackage{bm}
\usepackage{mathrsfs}
\usepackage{extarrows}
\usepackage{amsthm}
\usepackage{cases}
\usepackage{subeqnarray}
\usepackage{float}
\usepackage{cite}
\usepackage{stfloats}
\usepackage[noend]{algpseudocode}
\usepackage{algorithmicx,algorithm}

\theoremstyle{remark}

\usepackage{subfigure}
\usepackage{soul}
\usepackage{color}
\usepackage{booktabs}
\usepackage{balance}
\usepackage{epstopdf}

\begin{document}
\bibliographystyle{IEEEtran}

\title{\Huge{Deep Learning for Beam-Management:\\State-of-the-Art, Opportunities and Challenges}}

\author{Ke Ma, Zhaocheng Wang$^*$, \IEEEmembership{Fellow,~IEEE}, Wenqiang Tian, \\
Sheng Chen, \IEEEmembership{Fellow,~IEEE}, Lajos Hanzo, \IEEEmembership{Life Fellow,~IEEE}
\vspace*{-6mm}
}

\maketitle

\begin{abstract}

Benefiting from huge bandwidth resources, millimeter-wave (mmWave) communications provide one of the most promising technologies for next-generation wireless networks.
To compensate for the high pathloss of mmWave signals, large-scale antenna arrays are required both at the base stations and user equipment to establish directional beamforming, where beam-management is adopted to acquire and track the optimal beam pair having the maximum received power.
Naturally, narrow beams are required for achieving high beamforming gain, but they impose enormous training overhead and high sensitivity to blockages.
As a remedy, deep learning (DL) may be harnessed for beam-management.
First, the current state-of-the-art is reviewed, followed by the associated challenges and future research opportunities.
We conclude by highlighting the associated DL design insights and novel beam-management mechanisms.

\end{abstract}

\IEEEpeerreviewmaketitle

\section{Introduction}\label{sec:intro}

According to Ericsson's mobility report \cite{ref1}, the worldwide total monthly mobile data traffic will increase by 30\% each year and reach 143 exabytes in 2026.
To meet the ultra-high data traffic requirement, the enhanced Mobile Broadband (eMBB) mode of the fifth-generation (5G) wireless network has been designed for supporting high-speed access for users in hot-spot areas.
Millimeter-wave (mmWave) communications, benefiting from abundant bandwidth resources spanning from 30GHz to 300GHz have the potential of supporting Gigabits-per-second data rates.

However, mmWave carriers suffer from higher pathloss than those of conventional low-frequency communication systems.
Fortunately, the short wavelength of mmWave signals allows more antennas to be integrated into both the base stations (BSs) and user equipment (UE). Therefore, large-scale antenna arrays can be used at the BS and UE sides to implement directional beamforming, so that the high pathloss can be compensated by the beamforming gain.
To provide seamless high-quality services, beam-management has to be adopted to acquire and track the optimal BS and UE beam pair having the maximum received power.

However, these narrow beams impose beam-management challenges. On the one hand, numerous narrow candidate beams have to be created for covering the whole angular space, but this imposes substantial beam-training overhead. As a further challenge, narrow beams are sensitive to blockages, making accurate beam-tracking more difficult.

Inspired by the stunning breakthroughs that deep learning (DL) has achieved in computer vision and natural language processing, DL has also been harnessed in wireless communications \cite{refAI1, refAI2}.
Compared to mathematical model based methods, DL enjoys a pair of key advantages.
Firstly, mathematical tools generally rely on idealized assumptions, such as the presence of pure additive white Gaussian noise, which may not be consistent with practical scenarios.
By contrast, DL adaptively learns the features of the channel in support of reliable beam-management \cite{ref2, ref3}.
Secondly, the parameters of DL models capture the high-dimensional features of the propagation scenario, such as blockage locations and shapes \cite{ref4} in support of reliable beam-management, as detailed in the paper.

Based on these motivations, we survey the state-of-the-art by pursuing three avenues.
Firstly, DL is utilized to extract the nonlinear features inherent in the angular domain for implementing super-resolution beam-prediction.
Secondly, the complex relationships between the UE's movement and the channel variations inspire DL-based low-overhead beam-tracking and predictive beam-switching.
Thirdly, low-frequency channel information can be exploited to infer environmental features based on DL for assisting mmWave beam-management.
Finally, future research challenges and opportunities are summarized, where we provide design insights in support of practical implementations, and envision novel beam-management mechanisms relying on DL.

\section{Why DL for Beam-Management?}

DL is well-known for its beneficial adaptive fitting capability to extract complex nonlinear environmental features in the typical scenarios shown in Fig.~1.

\subsection{AoA/AoD vs. UE Position}

In practical scenarios, beam-management is faced with various sources of nonlinearities. For ease of handling, conventional mathematical methods usually simplify the real-world scenarios by ignoring these nonlinear factors.
By contrast, DL is capable of accurately modelling these complex nonlinear relationships and exploiting them to facilitate efficient beam-management.

\begin{figure*}[tp!]
\vspace{-4mm}
\begin{center}
\includegraphics[width=0.7\textwidth]{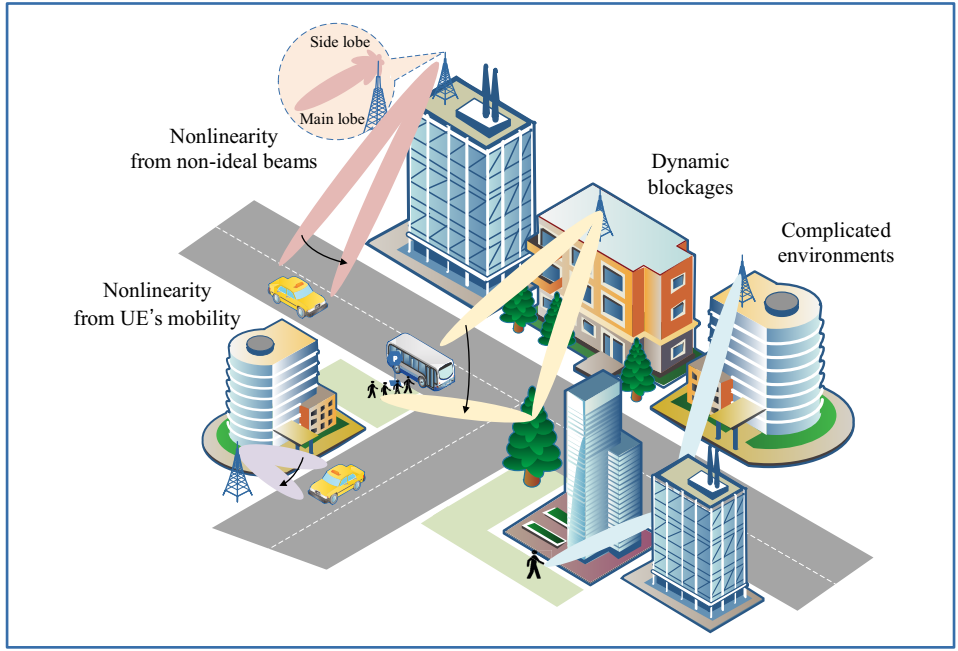}
\end{center}
\vspace{-3mm}
\caption{Illustration of typical scenarios applying DL in beam-management.}
\label{Fig1}
\vspace{-3mm}
\end{figure*}

Let us consider an example.
MmWave beams are usually generated by analog phase shifters, whose unit-amplitude constraint imposes nonlinearity on the beam shapes, which is termed as the channel's power leakage.
As it transpires from Fig.~1, achieving high beamforming gain requires extremely accurate angular alignment.
In case of misalignment, part of the beam's power leaks to side lobes, but this property allows us to estimate the angle-of-arrival (AoA) and angle-of-departure (AoD) of the dominant path by exploiting the signals received during beam-training.
Based on this observation, Qi \emph{et~al.} \cite{ref2} proposed to measure only a limited subset of candidate beams and use DL to predict the optimal beam direction for reducing the beam-training overhead.

Another challenge in beam-management arises from the UE's mobility.
Consider a line-of-sight (LOS) mobile scenario shown in Fig.~1. As the UE moves closer to the BS, the angular variation of the LOS path would become faster as a function of their distance and vice versa.
Moreover, the velocity and direction of UEs is time-variant.
This nonlinear relationship between the UE's movement and angular variation makes it hard to accurately track the beam directions, especially for high-speed scenarios.
To tackle this issue, Liu \emph{et~al.} \cite{ref3} proposed to use deep reinforcement learning (DRL) to model the nonlinear variations of the LOS angle, which outperformed the extended Kalman filter based tracking.

\subsection{High-Dimensional Feature Space}

In wireless environments, numerous scatterers having diverse locations, sizes and shapes appear, which form high-dimensional feature spaces. They jointly determine the optimal beam direction.

DL is capable of accurately modelling these complex scattering features.
Specifically, the AoA and AoD implicitly characterize the interaction between the transmitted signals and the propagation environment around the BS and UE.
Therefore, the training data collected allows DL to accurately model the complex relationships between the optimal beam direction and the specific environments.
For example, Chen \emph{et~al.} \cite{ref4} exploited the channel state information (CSI) dependence between a pair of neighboring BSs encountering joint scatterers, and proposed to use DL to infer the optimal beam direction of the target BS according to the source BS's CSI. They also demonstrated that the total training overhead of the system relying on multiple BSs was significantly decreased.

\subsection{Adaptive Fitting Capability}

Due to the mobility of users and scatterers, blockages tend to dynamically appear and disappear. As shown in Fig.~1, when a bus stops at the station and blocks the LOS path, the beam must be switched to another direction aligned with the strongest non-line-of-sight (NLOS) path. In this situation, beam-management has to sense the environmental variation and determine the new optimal beam direction. Conventional beam-management methods usually detect the blockage by using a received power threshold and then they perform perform re-sweeping, which leads to excessive overhead.

The main advantage of integrating DL into beam-management is the ability to adapt to the dynamic fluctuations of the environment. When the DL model is well trained, continued online training can be activated, which keeps collecting training data and updating the model parameters according to the environmental variations.
Alkhateeb \emph{et~al.} \cite{ref5} investigated online training at a bus station subjected to periodical blockages. After training based on both LOS and NLOS data, the instantaneous blockages only impose slight degradation of the predicted beamforming gain, which validates the robustness of DL to dynamic environmental fluctuations.

\section{State-of-the-Art}

Applying DL in beam-management offers the following research routes.
Firstly, it can be utilized for modelling both the angular- and time-domain features, which inspires both super-resolution beam-prediction and beam-tracking together with predictive beam-switching.
Secondly, DL may be adopted for extracting high-dimensional environmental features based on DL for assisting beam-management.

\subsection{Super-Resolution Beam-Prediction}

\begin{figure*}[tp!]
\vspace{-4mm}
\begin{center}
\includegraphics[width=0.9\textwidth]{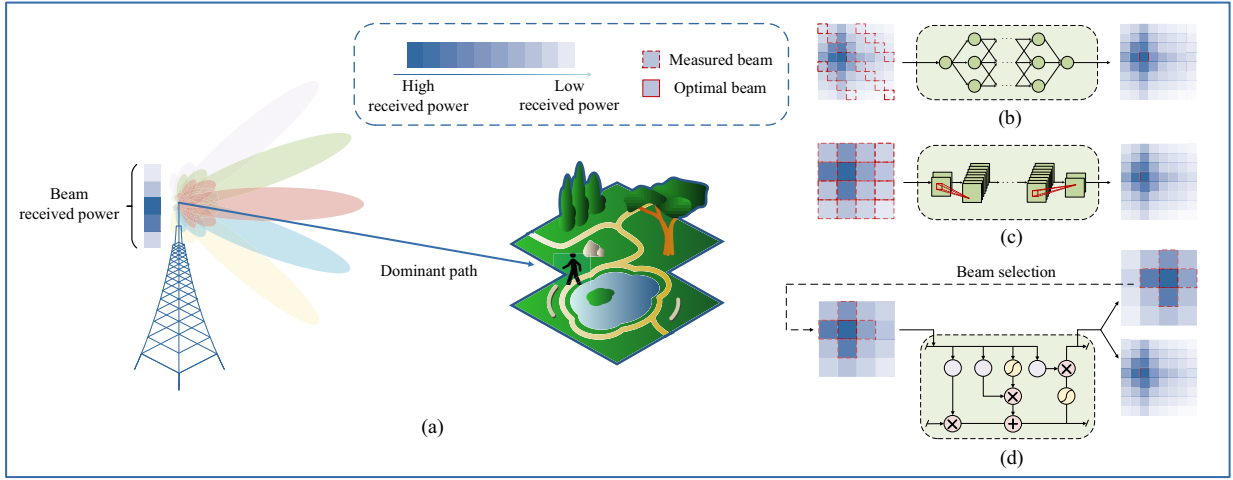}
\end{center}
\vspace{-3mm}
\caption{(a) An example of the channel's power leakage; (b)$\sim$(d) illustration of various super-resolution beam-prediction schemes: (b) DNN based prediction using fixed angular-spacing beams; (c) CNN based prediction using wide beams; (d) LSTM based prediction using partial high-SNR wide beams.}
\label{Fig2}
\vspace{-3mm}
\end{figure*}

The excessive overhead of beam-training poses a significant challenge in mmWave communications. For example, for a mmWave system considering 64 beams at the BS and 8 beams at the UE, 512 measurements are required for finding the optimal beam pair.
To reduce this excessive complexity, DL may be used for predicting the optimal high-resolution beam by using low-resolution beam-search at a low overhead.

An intuitive technique of reducing the beam-training overhead is to only consider the beams having a fixed angular spacing and then use their received signals to predict the optimal beam \cite{ref2}, as shown in Fig.~2(b).
Specifically, a fully-connected (FC) deep neural network (DNN) is established for modelling the angular features, followed by a softmax function that converts the output into probabilities.
Then the maximum-probability beam is selected.
However, its performance may degrade due to the low signal-to-noise ratio (SNR) when the true AoA and AoD of the dominant path does not fall exactly in the middle of the main lobe of any candidate beam.

To fully cover the 360$^\circ$ angular space, Echigo \emph{et~al.} \cite{ref6} proposed a wide-beam based optimal narrow-beam prediction scheme. The concept of wide and narrow beams arises from a twin-level beam-search philosophy based on a hierarchical multi-resolution codebook, where the first-level search aims for finding the optimal wide beam. Then the second level search confirms the specific optimal narrow-beam direction within the range of the selected wide beam. As seen in Fig.~2(c), the wide and narrow beams can be naturally regarded as low-resolution and high-resolution beams. Consequently, Echigo \emph{et~al.} \cite{ref6} compared super-resolution beam-prediction to a super-resolution image recovery problem, and adopted a convolutional neural network (CNN) for the associated prediction.

Furthermore, as shown in Fig.~2(a), the leaked power of the beams which are angularly far from the AoA and AoD of the dominant path is low. Hence it is difficult to extract useful information from the corresponding received signals due to their low SNRs. Therefore, Ma \emph{et~al.} \cite{ref7} proposed to consider only the subset of wide beams having high SNRs and use their received signals to predict the optimal narrow beam. This way the wide-beam-training overhead was further reduced. Specifically, a long-short term memory (LSTM) network was constructed for modelling the temporal AoA and AoD variations, as seen in Fig.~2(d). At the $t$-th time slot, the proposed network predicts not only the $t$-th optimal narrow beam, but also the $(t+1)$-st optimal wide beam for selecting the neighboring high-SNR wide beams in the $(t+1)$-st wide-beam-training.

Figure 3 compares the above DL-based schemes and the conventional noise-free narrow-beam estimation scheme based on wide beams in terms of their effective spectral efficiency, which takes the beam-training overhead into consideration.
It is clear that DL-based schemes outperform their conventional counterparts, because DL models the nonlinear power leakage phenomenon more accurately.
Furthermore, the partial-search based high-SNR wide-beam assisted scheme achieves higher spectral efficiency due to its lower training overhead, especially for short beam-training periods.

\begin{figure}[tp!]
\vspace{-1mm}
\begin{center}
\includegraphics[width=0.45\textwidth]{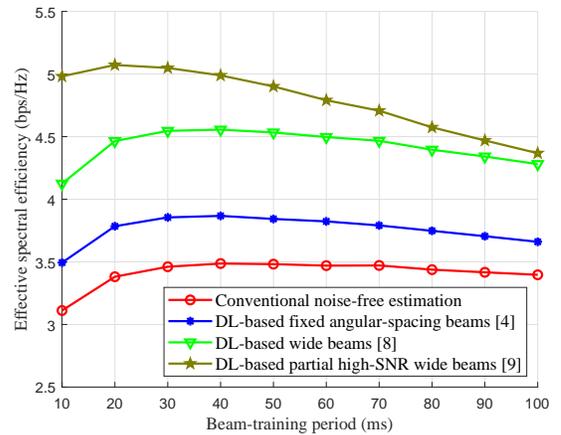}
\end{center}
\vspace{-5mm}
\caption{Effective spectral efficiency as function of beam-training period for various schemes.}
\label{Fig3}
\end{figure}

\subsection{Beam-Tracking and Predictive Beam-Switching}

\begin{figure*}[tp!]
\vspace{-4mm}
\begin{center}
\includegraphics[width=0.86\textwidth]{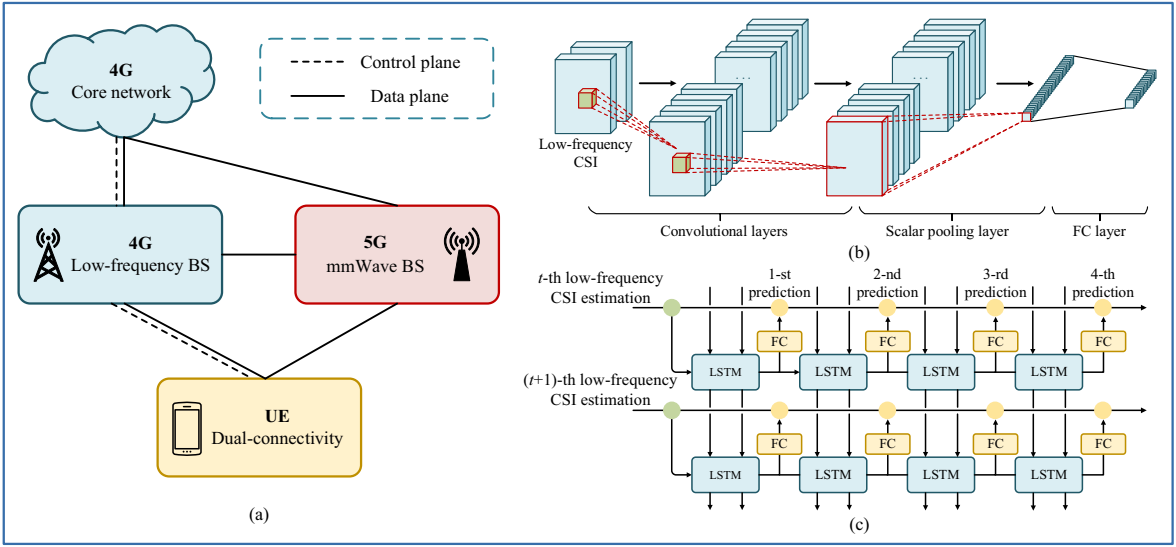}
\end{center}
\vspace{-3mm}
\caption{(a) NSA architecture; (b) CNN based mmWave beam-prediction using low-frequency CSI; (c) cascaded LSTM based mmWave beam-tracking using low-frequency CSI.}
\label{Fig4}
\vspace{-3mm}
\end{figure*}

A crucial challenge for beam-management is to tackle rapidly time-varying channels, which requires frequent beam-training for maintaining seamless high-quality services.
Fortunately, the predictability of the UE's trajectory assists in reducing the beam-management overhead.

Lim \emph{et~al.} \cite{ref8} combined the conventional beam-tracking scheme with DL.
At each beam-tracking action, firstly the LSTM network is exploited for fusing the previous CSI estimates and sensor measurements to extract the UE's movement features for predicting the \emph{a-priori} AoA distribution.
Based on this predicted distribution, the future beams are predicted and then measured for creating the more accurate \emph{a-posteriori} estimate from the \emph{a-priori} CSI estimate by using sequential Bayesian estimation.
This scheme intrinsically integrates the expert knowledge of classical mathematical models with the adaptive fitting capability of DL, and thus achieves lower tracking errors than the conventional DL-based scheme that only relies on training data.

The UE's ever-changing mobility is another key issue influencing beam-tracking.
Explicitly, when the velocity is high, the tracking range has to be increased and vice versa. Therefore, Zhang \emph{et~al.} \cite{ref9} proposed to use DRL for seamlessly adapting to the tracking range.
Concretely, the tracking action is jointly determined by the initial beam index and the size of the tracking beam subset.
The proposed deep Q-learning model interacts with the environment, adjusting the tracking range according to the effective throughput reward that takes both the beamforming gain and the tracking overhead into account.

As a further advance, a predictive beam-switching scheme based on DL was proposed in \cite{ref6}. After each beam-training action, the LSTM network is utilized to predict the optimal beam in the middle of the current and next beam-training instants according to the previous received beam-training signals, which halves the training overhead. Echigo \emph{et~al.} \cite{ref6} additionally integrated predictive beam-switching with super-resolution prediction for further reducing the training overhead.

\subsection{Low-Frequency Information Assisted mmWave Beam-Prediction}

\begin{figure*}[tp!]
\vspace{-4mm}
\begin{center}
\includegraphics[width=1.0\textwidth]{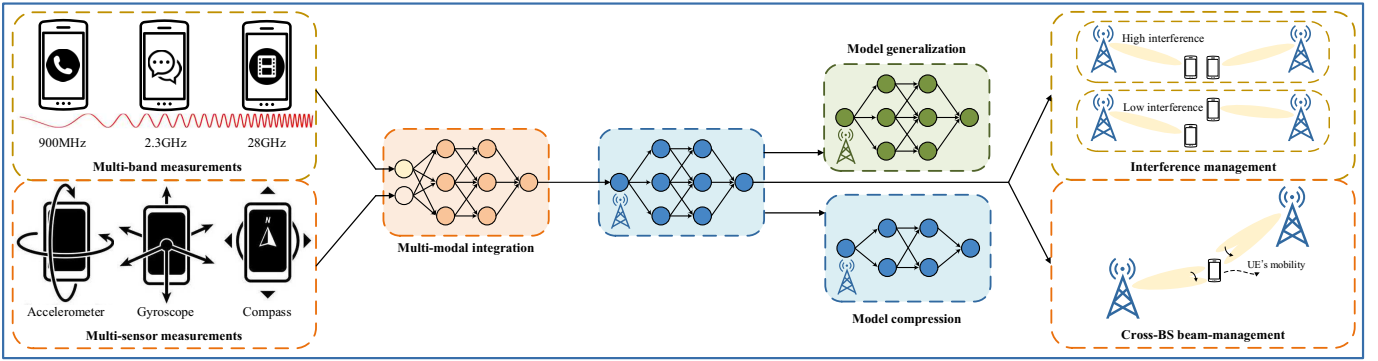}
\end{center}
\vspace{-3mm}
\caption{Summary of research challenges and opportunities for DL assisted beam-management.}
\label{Fig5}
\vspace{-3mm}
\end{figure*}

The non-standalone (NSA) architecture that integrates existing low-frequency fourth-generation (4G) facilities with mmWave equipment is one of the fundamental 5G architectures.
As shown in Fig.~4(a), in the NSA architecture, the control function relies on the 4G control plane, while the mmWave BSs are dedicated to data rate enhancement, where dual-connectivity relying on both low-frequency and mmWave links is supported at the UE.
Usually, mmWave antennas are used also by the low-frequency BSs to reduce hardware cost.
In this scenario, the low-frequency and mmWave links perceive having similar propagation environments, and thus the mmWave channel enjoys analogous AoA and AoD features to the low-frequency counterpart.
Therefore, DL may be exploited for revealing the complex relationships between the low-frequency and the mmWave channels by exploiting their shared environmental features.
This allows the prediction of the optimal mmWave beam by relying on low-frequency channel information.

Given the high power consumption of mmWave devices, the mmWave link is only activated upon requiring high data rates, where the low-frequency CSI can be used for predicting the optimal mmWave beam during the initial access. Alrabeiah \emph{et~al.} \cite{ref10} conceived a fully-connected (FC) DNN for the associated prediction, where the softmax function converts the output into probabilities. Based on these probabilities, Alrabeiah \emph{et~al.} \cite{ref10} proposed to measure a limited subset of candidate beams having the highest probabilities, and to select the specific beam having the maximum received power as the optimal one, which achieved more accurate beam-alignment in exchange for a modest extra overhead.

Practically, the length of the low-frequency CSI is dynamically varying, since different UEs have different numbers of antennas and subcarriers. To process this variable-length CSI-input, Ma \emph{et~al.} \cite{ref11} proposed a dedicated CNN for the associated prediction, as shown in Fig.~4(b). Specifically, because convolution operations do not require a fixed input-length, firstly convolutional layers are utilized to extract some hidden features from the low-frequency CSI. Then, the FC layer is adopted to match the size of the extracted features to the specific number of candidate beams. Since the FC layer requires fixed-length input, a scalar pooling layer is introduced after the convolutional layers, which downsamples the extracted features to a scalar, thus guaranteing the efficient support of variable-length CSIs.

Instead of using the instantaneous CSI, Ma \emph{et~al.} \cite{ref12} proposed to harness the previous low-frequency CSI sequence to track the optimal mmWave beam, since the low-frequency CSI is periodically estimated for supporting the operation of the low-frequency link itself. However, due to having narrow beam widths, mmWave beam-prediction has to be updated more frequently than low-frequency CSI estimation. To address this issue, the cascaded LSTM philosophy of Fig.~4(c) was proposed for predicting the optimal mmWave beam at multiple uniformly distributed instants between two low-frequency CSI estimations.

\section{Research Challenges and Opportunities}

Although DL has achieved encouraging performance gains in beam-management, there are numerous challenges for further study, as summarized in Fig.~5 at a glance.

\subsection{DL Design}

\textbf{Model compression}: Given the UE's mobility, beam-alignment is usually refreshed every dozen of milliseconds or so.
Therefore, DL must be completed within this period for ensuring prompt beam-alignment.
To meet this high complexity requirement, DL has to be deployed at the BS, which has powerful computational capability.
Nonetheless, reducing the numerous model parameters must be further investigated.
For example, the popular teacher-student model can be exploited, where the powerful teacher model firstly extracts the associated complex nonlinear features, and then the smaller student model is adopted to fit the teacher model, so that the overall model size is reduced.

\textbf{Model generalization}: The optimal beam angle depends on the propagation environment. Naturally, each BS and UE best collects its own training data comprising the specific environmental features for optimizing the local model, but collecting huge amounts of training data imposes excessive cost. Fortunately, many of these environmental factors share similar properties (such as the LOS path), while others vary (such as scatterers) for different environments. Therefore, transfer-learning can be adopted for applying the knowledge extracted from one environment to another one, so that satisfactory performance can be achieved, despite using only limited training sets. In this spirit, Rezaie \emph{et~al.} \cite{ref13} proposed to freeze the output layer in beam-prediction and to only adjust the hidden layers for adapting to various indoor environments. This demonstrated significant prediction accuracy enhancement under small datasets. As a further emerging technology, meta-learning can be adopted for further enhancing the model's generalization capability.

\subsection{Beam-Management Mechanism}

\textbf{Multi-modal integration}:
Next-generation terminals will support multiple radio frequency (RF) bands, such as 900MHz for voice calls, 2.3GHz for basic data services, 28GHz and beyond for high-speed access, where different bands interact with the environment in different manners.
For example, low-frequency signals are more likely to diffract over a pedestrian, while mmWave signals might be blocked.
Therefore, the fusion of various bands assists in characterising the propagation environment more precisely, in order to facilitate accurate beam-alignment.
Moreover, the UE's sensor measurements can provide further efficient auxiliary information for beam-management, such as the gyroscope for detecting rotations, which can be used to calibrate the UE's beam direction.
For diverse input sizes and formats, having a robust DL-based multi-modal integration framework is vital for adaptively fusing the features extracted from each input.

\textbf{Interference management}: In contrast to the omni-directional antenna used in low-frequency links, highly directional beams make the inter-cell interference erratic, since it mainly depends on the direction of the beam used by the neighboring cells.
Specifically, the beam impinging from a neighboring cell pointing towards the local UE imposes severe interference.
However, the UE's mobility leads to frequent beam-switching, hence requiring agile interference management.
Based on the smoothly evolving trajectory of the UE's movement, recurrent neural networks such as LSTM can foresee the forthcoming interference occurrences and activate interference cancellation in advance.

\textbf{Cross-BS beam-management}: In mobility management, the prompt switchings of both the optimal BS and beam are crucial for providing seamless high-quality services. Powerful DL can be used for establishing a cross-BS beam-management framework, where the BS selection and beam selection are performed jointly for attaining ultra-low handover latency. In this framework, DRL provides a promising tool for modelling diverse propagation environments in support of the interaction between the serving BS and neighboring BSs, so that both the BS selection and beam selection can be adaptively adjusted for maintaining the service quality.


\section{Conclusions}

We commenced by elaborating on the motivation of applying DL in beam-management. Firstly, DL is eminently suitable for extracting the complex nonlinear environmental features encountered. Secondly, the adaptive fitting capability of DL enables near-real-time fine-tuning required by the environmental fluctuations. Based on these motivations, DL was proposed for extracting both the nonlinear angular- and time-domain features, hence inspiring super-resolution beam-prediction and beam-tracking along with predictive beam-switching. Furthermore, low-frequency channel information can be adopted to infer the propagation features for assisting mmWave beam-management. Finally, we have provided DL design insights for the practical implementation of novel beam-management mechanisms.

\section*{Acknowledgment}
This work was supported in part by the National Key R\&D Program of China under Grant 2018YFB1801102 and in part by the National Natural Science Foundation of China (Grant No. 61871253). This work was also supported by OPPO Research Fund \emph{(Corresponding Author: Zhaocheng Wang)}.

{

}

\begin{IEEEbiographynophoto}{Ke Ma}
received his B.S. degree from Tsinghua University in 2019, where he is currently working toward the Ph.D. degree with the Department of Electronic Engineering, Tsinghua University. His current research interests include mmWave communications and intelligent communications.
\end{IEEEbiographynophoto}

\begin{IEEEbiographynophoto}{Zhaocheng Wang}
(F'21) received his B.S., M.S., and Ph.D. degrees from Tsinghua University in 1991, 1993, and 1996, respectively. From 1996 to 1997, he was a Post Doctoral Fellow with Nanyang Technological University, Singapore. From 1997 to 2009, he was a Research Engineer/Senior Engineer with OKI Techno Centre Pte. Ltd., Singapore. From 1999 to 2009, he was a Senior Engineer/Principal Engineer with Sony Deutschland GmbH, Germany. Since 2009, he has been a Professor with Department of Electronic Engineering, Tsinghua University. He was a recipient of IEEE Scott Helt Memorial Award, IET Premium Award, IEEE ComSoc Asia-Pacific Outstanding Paper Award and IEEE ComSoc Leonard G. Abraham Prize.
\end{IEEEbiographynophoto}

\begin{IEEEbiographynophoto}{Wenqiang Tian}
received his B.S. degree from Fudan University in 2010 and his Ph.D. degree from University of Chinese Academy of Sciences in 2015. Now, he is a senior Standardization Researcher of Guangdong OPPO Mobile Telecommunications Corp., Ltd. He has participated in the 5G standardization work and focused on physical layer design in 3GPP R15 and R16.
\end{IEEEbiographynophoto}

\begin{IEEEbiographynophoto}{Sheng Chen}
(F'08) received his B.Eng. degree in control engineering from East China Petroleum Institute, China, in 1982, and his Ph.D. degree in control engineering from City University, U.K., in 1986, and the Doctor of Sciences (D.Sc.) degree from the University of Southampton, U.K., in 2005. From 1986 to 1999, He held research and academic appointments with the Universities of Sheffield, Edinburgh and Portsmouth, all in U.K. Since 1999, he has been with the School of Electronics and Computer Science, University of Southampton, U.K., where he holds
the post of a Professor in intelligent systems and signal processing. He is a Fellow of the Royal Academy of Engineering (FREng), of the Asia-Pacific Artificial Intelligence Association and of the IET.
\end{IEEEbiographynophoto}

\begin{IEEEbiographynophoto}{Lajos Hanzo}
(http://www-mobile.ecs.soton.ac.uk, https://en.wiki-pedia.org/wiki/Lajos$\underline{~}$Hanzo) (F'04) received his Master degree and Doctorate in 1976 and 1983, respectively from the Technical University (TU) of Budapest. He was also awarded the Doctor of Sciences (D.Sc.) degree by the University of Southampton (2004) and Honorary Doctorates by the TU of Budapest (2009) and by the University of Edinburgh (2015).  He is a Foreign Member of the Hungarian Academy of Sciences and a former Editor-in-Chief of the IEEE Press. He has served several terms as Governor of both IEEE ComSoc and of VTS. He is also a Fellow of the Royal Academy of Engineering (FREng), of the IET and of EURASIP. He is the recipient of the 2022 Eric Sumner Field Award.
\end{IEEEbiographynophoto}


\begin{thebibliography}{10}

\bibitem{ref1} 
\emph{Ericsson's mobility report}, Accessed: Nov. 1, 2021. [Online]. Available: https://www.ericsson.com/en/reports-and-papers/mobility-report.


\bibitem{refAI1}
H.~Huang, \emph{et al.}, ``Deep learning for physical-layer 5G wireless techniques: opportunities, challenges and solutions,'' \emph{IEEE Wireless Commun.}, vol.~27, no.~1, pp.~214--222, Feb.~2020.

\bibitem{refAI2}
L.~Dai, \emph{et al.}, ``Deep learning for wireless communications: an emerging interdisciplinary paradigm,'' \emph{IEEE Wireless Commun.}, vol.~27, no.~4, pp.~133--139, Aug.~2020.

\bibitem{ref2} 
C.~Qi, Y.~Wang, and G.~Y~ Li, ``Deep learning for beam training in millimeter wave massive MIMO systems,'' \emph{IEEE Trans. Wireless Commun.}, early access, doi: 10.1109/TWC.2020.3024279.

\bibitem{ref3} 
Y.~Liu, Z.~Jiang, S.~Zhang, and S.~Xu, ``Deep reinforcement learning based beam tracking for low-latency services in vehicular networks,'' in \emph{Proc. ICC 2020}, Jun.~7--11, 2020, pp.~1--7.

\bibitem{ref4} 
S.~Chen, \emph{et al.}, ``Learning-based remote channel inference: feasibility analysis and case study,'' \emph{IEEE Trans. Wireless Commun.}, vol.~18, no.~7, pp.~3554--3568, Jul.~2019.

\bibitem{ref5} 
A.~Alkhateeb, \emph{et al.}, ``Deep learning coordinated beamforming for highly-mobile millimeter wave systems,''  \emph{IEEE Access}, vol.~6, pp.~37328--37348, Jun. 2018.

\bibitem{ref6} 
H.~Echigo, \emph{et al.}, ``A deep learning-based low overhead beam selection in mmWave communications,'' \emph{IEEE Trans. Veh. Tech.}, vol.~70, no.~1, pp.~682--691, Jan.~2021.

\bibitem{ref7} 
K.~Ma, \emph{et al.}, ``Deep learning assisted calibrated beam training for millimeter-wave communication systems,'' \emph{IEEE Trans. Commun.}, vol.~69, no.~10, pp.~6706--6721, Oct.~2021.

\bibitem{ref8} 
S.~H.~Lim, \emph{et al.}, ``Deep learning-based beam tracking for millimeter-wave communications under mobility,'' \emph{IEEE Trans. Commun.}, early access, doi: 10.1109/TCOMM.2021.3107526.

\bibitem{ref9} 
J.~Zhang, \emph{et al.}, ``Intelligent interactive beam training for millimeter wave communications,'' \emph{IEEE Trans. Wireless Commun.}, vol.~20, no.~3, pp.~2034--2048, Mar. 2021.

\bibitem{ref10} 
M.~Alrabeiah and A.~Alkhateeb, ``Deep learning for mmWave beam and blockage prediction using sub-6 GHz channels,'' \emph{IEEE Trans. Commun.}, vol.~68, no.~9, pp.~5504--5518, Sep. 2020.

\bibitem{ref11} 
K.~Ma, P.~Zhao, and Z.~Wang, ``Deep learning assisted robust mmWave beam prediction using low-frequency information,'' to appear in \emph{China Commun.}

\bibitem{ref12} 
K.~Ma, \emph{et al.}, ``Deep learning assisted mmWave beam prediction with prior low-frequency information,'' in \emph{Proc. ICC 2021}, Jun.~14--23, 2021, pp.~1--6.

\bibitem{ref13} 
S.~Rezaie, \emph{et al.}, ``Deep transfer learning for location-aware millimeter wave beam selection,'' \emph{IEEE Commun. Lett.}, vol.~25, no.~9, pp.~2963--2967, Sep. 2021.

\end{thebibliography}
\end{document}